\documentclass[10pt,amsmath,twocolumn,aps,prl]{revtex4-1}

\usepackage{graphicx}
\usepackage{amssymb}
\usepackage{amsmath}
\usepackage{textcomp}

\begin{document}

\title{Petermann-factor limited sensing near an exceptional point}
\author{Heming Wang$^{1,\dagger}$, Yu-Hung Lai$^{1,\dagger}$, Zhiquan Yuan$^{1,\dagger}$, Myoung-Gyun Suh$^{1}$ and Kerry Vahala$^{1, *}$\\
$^1$T. J. Watson Laboratory of Applied Physics, California Institute of Technology, Pasadena, California 91125, USA.\\
$^\dagger$These authors contributed equally.\\
$^*$Corresponding author: vahala@caltech.edu
}

\maketitle

\noindent {\bf 
Non-Hermitian Hamiltonians \cite{NonHermitianHamiltonian1,NonHermitianHamiltonian2} describing open systems can feature singularities called exceptional points (EPs) \cite{NonHermitianOptics1,NonHermitianOptics2,NonHermitianOptics3}. Resonant frequencies become strongly dependent on externally applied perturbations near an EP which has given rise to the concept of EP-enhanced sensing in photonics \cite{EPSensingP1, EPSensingP2, EPSensingP3, EPSensingP4}. However, while increased sensor responsivity has been demonstrated \cite{EPSensingE1,EPSensingE2,Gyro}, it is not known if this class of sensor results in improved signal-to-noise performance \cite{EPLimit1,EPLimit2,EPLimit3,EPLimit4,EPTechnicalNoise}. Here, enhanced responsivity of a laser gyroscope caused by operation near an EP is shown to be exactly compensated by increasing sensor noise in the form of linewidth broadening. The noise, of fundamental origin, increases according to the Petermann factor \cite{PF1,PF2},  because the mode spectrum loses the oft-assumed property of orthogonality. This occurs as system eigenvectors coalesce near the EP and a biorthogonal analysis confirms experimental observations. Besides its importance to the physics of microcavities and non-Hermitian photonics, this is the first time that fundamental sensitivity limits have been quantified in an EP sensor.
}

EPs have been experimentally realized in several systems \cite{EP1,EP2,EP3} and applied to demonstrate non-reciprocal transmission \cite{EPNonRecip1,EPNonRecip2,EPNonRecip3}, lasing dynamics control \cite{EPLasing1,EPLasing2,EPLasing3,EPLasing4} and improved optical-based sensors \cite{EPSensingP1, EPSensingP2, EPSensingP3, EPSensingP4}. 
The connection of the Petermann factor  \cite{PF1,PF2,ExcessNoise1,ExcessNoise2,ExcessNoiseEP1} to EPs  was considered in theoretical studies of microresonators \cite{PFResonator1,PFResonator2}. However, despite continued theoretical interest  \cite{PFResonator3,PFResonator4} including the development of new techniques for determination of linewidth in general laser systems \cite{GeneralLinewidth1}, the observation of Petermann linewidth broadening near exceptional points was reported only recently by the Yang group in a phonon laser system \cite{PhononLaserLinewidth}. And the link between Petermann-factor-induced noise and EP sensor performance has not been considered. 

Recently, strong sensing improvement near an EP was reported in gyroscopes operating near an EP \cite{Gyro}. Here it is shown that mode non-orthogonality induced by the EP severely limits the benefit of this improvement. Indeed, analysis and measurement confirm near perfect cancellation of signal improvement by increasing noise so that gyroscope signal-to-noise ratio (SNR) and hence sensitivity is not improved by operation near the EP.  The impact of mode non-orthogonality on laser noise is analyzed using the Petermann factor and compared with measurements. These results are further confirmed using an Adler phase locking equation approach \cite{Adler} which is also applied to analyze the combined effect of dissipative and conservative coupling on the system.

\begin{figure*}
\centering
\includegraphics[width=170mm]{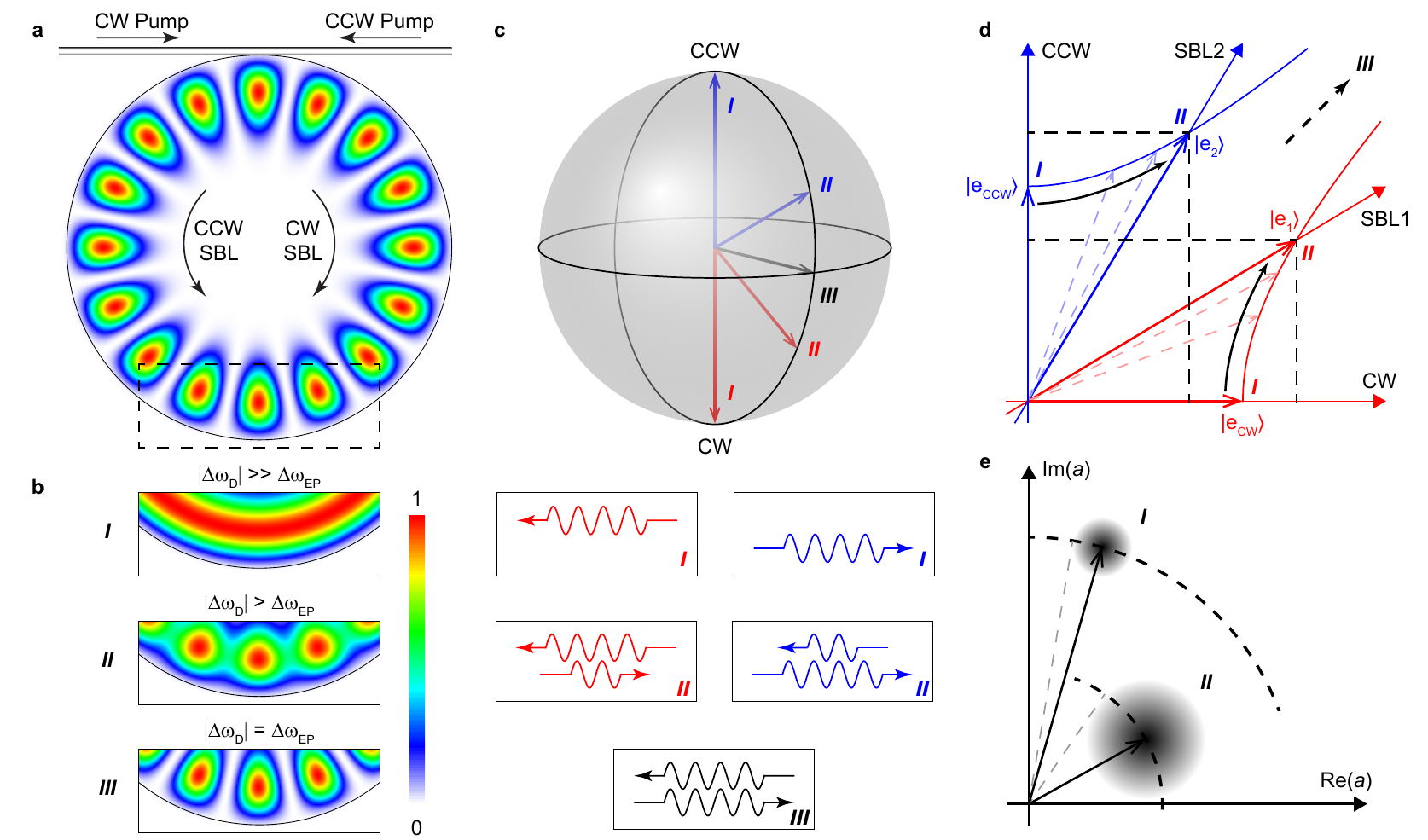}
\caption{{\bf Brillouin laser linewidth enhancement near an exceptional point.}
{\bf a}, Diagram of whispering-gallery mode resonator with the energy distribution of an eigenmode superimposed.
A portion of the resonator is outlined corresponding to state {\it III} in panel {\bf b}. Optical pumps on the coupling waveguide and whispering-gallery SBL modes are indicated by arrows.
{\bf b}, Mode energy distributions for three different states: far from EP ({state \it I}) the eigenmode is a traveling cw or ccw wave; near EP ({state \it II}) the eigenmode is a hybrid of cw and ccw waves; at EP ({state \it III}) the eigenmode is a standing wave.
{\bf c}, Bloch sphere showing the eigenstates for cases {\it I}, {\it II} and {\it III} with corresponding cw and ccw composition.
{\bf d}, Illustration of the cw-ccw and SBL1-SBL2 coordinate systems. Unit vectors for states {\it I} and {\it II} are shown on each axis. As the system is steered towards the EP, the SBL axes move toward each other so that unit vectors along the SBL axes lengthen as described by the two hyperbolas. This is illustrated by decomposing a unit vector of the non-orthogonal SBL coordinate system using the orthogonal cw-ccw coordinates [e.g., $(5/4,3/4)^T$ and $(3/4,5/4)^T$ for state {\it II}]. Consequently, the field amplitude is effectively shortened in the SBL basis.
{\bf e}, Phasor representation of the complex amplitude of a lasing mode for states {\it I} and {\it II} provides an interpretation of linewidth enhancement. 
Phasor length is shortened and noise is enhanced as the system is steered to the EP, leading to an increased phasor angle diffusion and laser linewidth enhancement (see Supplementary Information).
}
\label{fig1}
\end{figure*}

The gyroscope uses a high-Q silica whispering gallery resonator \cite{WedgeResonator} in a ring-laser configuration \cite{Gyro0}. As illustrated in Fig. \ref{fig1}a, optical pumping of clockwise (cw) and counter-clockwise (ccw) directions on the same whispering-gallery mode index induces laser action through the Brillouin process. On account of the Brillouin phase matching condition, these stimulated Brillouin laser (SBL) waves propagate in a direction opposite to their corresponding pump waves \cite{SBL1}. Dissipative backscattering couples the SBLs and above threshold the following Hamiltonian governs their motion \cite{Gyro}:
\begin{equation}
\label{H}
H=
\left(
\begin{matrix}
\omega_\mathrm{cw} & i\Delta\omega_\mathrm{EP}/2\\
i\Delta\omega_\mathrm{EP}/2 & \omega_\mathrm{ccw}
\end{matrix}
\right)
\end{equation}
where $H$ describes the dynamics via $id\Psi/dt=H\Psi$ and $\Psi=(a_\mathrm{cw},a_\mathrm{ccw})^T$ is the column vector of SBL mode amplitudes (square of norm is photon number). Also, $\Delta\omega_\mathrm{EP}$ is a non-Hermitian term related to the coupling rate between the two SBL modes and $\omega_\mathrm{cw}$ ($\omega_\mathrm{ccw}$) is the active-cavity resonance angular frequency of the cw (ccw) SBL mode above laser threshold. The dependence of $\omega_\mathrm{cw}$, $\omega_\mathrm{ccw}$ and $\Delta\omega_\mathrm{EP}$ on other system parameters, most notably the angular rotation rate and the optical pumping frequencies, has been suppressed for clarity. 

A class of EP sensors operate by measuring the frequency difference of the two system eigenmodes. This difference is readily calculated from Eq. (\ref{H}) as $\Delta\omega_\mathrm{S}= \sqrt{\Delta\omega_\mathrm{D}^2-\Delta\omega_\mathrm{EP}^2}$ where $\Delta\omega_\mathrm{D}\equiv\omega_\mathrm{ccw}-\omega_\mathrm{cw}$ is the resonance frequency difference and $\Delta\omega_\mathrm{EP}$ is the critical value of $\Delta\omega_\mathrm{D}$ at which the system is biased at the EP. As illustrated in Fig. \ref{fig1}b,c the vector composition of the SBL modes strongly depends upon the system proximity to the EP. For $|\Delta\omega_\mathrm{D}| \gg \Delta\omega_\mathrm{EP}$ the SBL modes (unit vectors) are orthogonal cw and ccw waves. However, closer to the EP the waves become admixtures of these states that are no longer orthogonal. At the EP, the two waves coalesce to a single state vector (a standing wave in the whispering gallery). Rotation of the gyroscope in state {\it II} in Fig. \ref{fig1} ($|\Delta\omega_\mathrm{D}|>\Delta\omega_\mathrm{EP}$) introduces a perturbation to $\Delta\omega_\mathrm{D}$ whose transduction into $\Delta\omega_\mathrm{S}$ is enhanced relative to the conventional Sagnac factor \cite{SagnacTheoryRLG}. This EP-induced signal-enhancement-factor (SEF) is given by \cite{Gyro},
\begin{equation}
\label{SEF}
\mathrm{SEF}=\left|\frac{\partial \Delta\omega_\mathrm{S}}{\partial\Delta\omega_\mathrm{D}}\right|^2=\frac{ \Delta\omega_\mathrm{D}^2}{\Delta\omega_\mathrm{D}^2-\Delta\omega_\mathrm{EP}^2}
\end{equation}
where SEF refers to the signal power (not amplitude) enhancement. This factor has recently been verified in the Brillouin ring laser gyroscope \cite{Gyro}. The control of $\Delta\omega_\mathrm{D}$ (and in turn $\Delta\omega_\mathrm{S}$) in that work and here is possible by tuning of the optical pumping frequencies and is introduced later.

$\Delta\omega_\mathrm{S}$ is measured as the beat frequency of the SBL laser signals upon photodetection and the SNR is set by the laser linewidth.  To understand its linewidth behavior a bi-orthogonal basis is used as described in the Supplementary Information.  As shown there and illustrated in Fig. \ref{fig1}d, the peculiar properties of non-orthogonal systems near the EP cause the unit vectors (optical modes) to be lengthened. This lengthening results in an effectively shorter laser field amplitude. Also, noise into the mode is increased as illustrated in Fig. \ref{fig1}e. Because the laser linewidth can be understood to result from diffusion of the phasor  in Fig. \ref{fig1}e, linewidth increases upon operation close to the EP. And the linewidth enhancement is given by the Petermann factor (see Supplementary Information),
\begin{equation}
\label{PF}
\mathrm{PF}=\frac{1}{2}\left(1+\frac{\mathrm{Tr}(H_0^\dag H_0)}{|\mathrm{Tr}(H_0^2)|}\right)=\frac{\Delta\omega_\mathrm{D}^2}{\Delta\omega_\mathrm{D}^2-\Delta\omega_\mathrm{EP}^2}
\end{equation}
where $\mathrm{Tr}$ is the matrix trace operation and $H_0=H-\mathrm{Tr}(H)/2$ is the traceless part of $H$. As derived in the Supplementary Information, the first part of this equation is a basis independent form and is valid for a general two-dimensional system. The second part is specific to the current SBL system. Inspection of Eq. (\ref{SEF}) and Eq. (\ref{PF}) shows that SEF $=$ PF. As a result the SNR is not expected to improve through operation near the EP when the system is fundamental-noise limited.

\begin{figure}
\centering
\includegraphics[width=8.5cm]{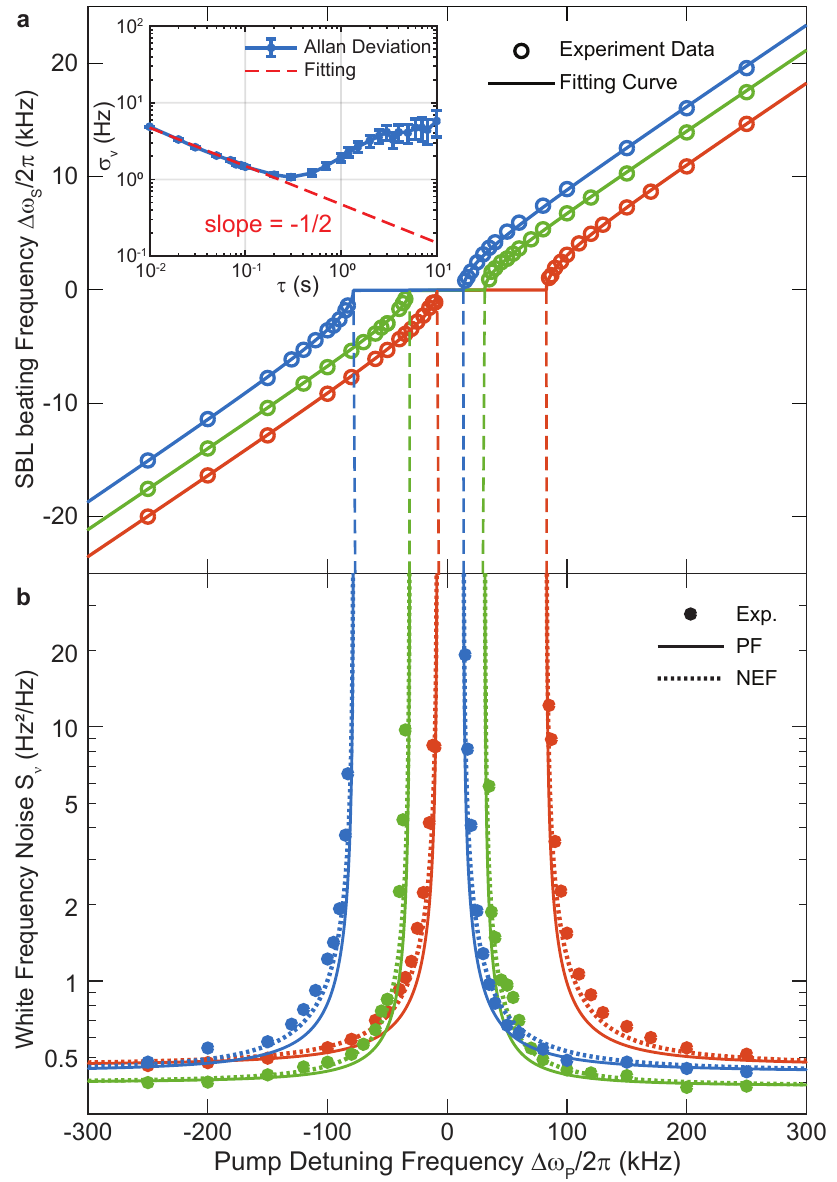}
\caption{\textbf{Measured linewidth enhancement of SBLs near the exceptional point.}
{\bf a}, Measured SBL beating frequency is plotted versus pump detuning for three distinct locking zones, corresponding SBL amplitude ratios $q$: 1.15 (blue), 1 (green), 0.85 (red).
Solid curves are theoretical fittings.
Inset is a typical Allan deviation of frequency $\sigma_\nu(\tau)$ versus gate time $\tau$. The short-term part is fitted with $\sqrt{S_\nu/(2\tau)}$ where $S_\nu$ is the one-sided power spectral density of the white frequency noise plotted in panel {\bf b}.
{\bf b}, Measured white frequency noise of the beating signal determined using the Allan deviation measurement. Data point color corresponds to same amplitude ratios used in panel {\bf a}. 
Petermann factor PF (solid lines) and the NEF (dashed lines) theoretical predictions use parameters obtained by fitting from panel {\bf a}.
}
\label{fig2}
\end{figure}

To verify the above predictions, the output of a single pump laser ($\sim$1553.3 nm) is divided into two branches that are coupled into cw and ccw directions of the resonator using a tapered fiber \cite{Taper1,Taper2}. Both pump powers are actively stabilized. The resonator is mounted in a sealed box and a thermo-electric cooler (TEC) controls the chip temperature which is monitored using a thermistor (fluctuations are held within 5 mK). Each pumping branch has its frequency controlled using acousto-optic modulators (AOMs). The ccw pump laser frequency is Pound-Drever-Hall (PDH) locked to one resonator mode and the cw pump laser can then be independently tuned by the AOM. This pump detuning frequency ($\Delta \omega_\mathrm{P}$) is therefore controlled to radio-frequency precision. It is used to precisely adjust $\Delta \omega_\mathrm{D}$ and in turn $\Delta\omega_\mathrm{S}$ as shown in three sets of measurements in Fig. \ref{fig2}a. Here, the photodetected SBL beat frequency $\Delta\omega_\mathrm{S}$ is measured using a frequency counter. The data sets are taken for three distinct SBL output amplitude ratios as discussed further below. A solid curve fitting is also presented using $\Delta\omega_\mathrm{S}=\pm\sqrt{\Delta \omega_\mathrm{D}^2-\Delta\omega_\mathrm{EP}^2}$, where $\Delta\omega_\mathrm{D}=\frac{\gamma/\Gamma}{1+\gamma/\Gamma}\Delta\omega_\mathrm{P}+\frac{1}{1+\gamma/\Gamma}\Delta\omega_{\mathrm{Kerr}}$ (see Supplementary information). Also, $\gamma$ is the photon decay rate, $\Gamma$ is the Brillouin gain bandwidth \cite{SBL1}, and $\Delta\omega_{\mathrm{Kerr}}$ is a Kerr effect correction that is explained below. As an aside, the data plot and theory show a frequency locking zone, the boundaries of which occur at the EP. 

The frequency counter data are also analyzed as an Allan deviation (Adev) measurement (Fig. \ref{fig2}a inset). The initial roll-off of the Adev features a slope of $-1/2$ corresponding to white frequency noise \cite{IEEEstandard}. This was also verified in separate measurements of the beat frequency using both an electrical spectrum analyzer and a fast Fourier transform. The slope of this region is fit to $\sqrt{S_\nu/(2\tau)}$ where $S_\nu$ is the one-sided spectral density of the white frequency noise. Adev measurement at each of the detuning points in Fig. \ref{fig2}a is used to infer the $S_\nu$ values that are plotted in Fig. \ref{fig2}b. There, a frequency noise enhancement is observed as the system is biased towards an EP. Also plotted is the Petermann factor noise enhancement (Eq. (3)). Aside from a slight discrepancy at intermediate detuning frequencies (analyzed further below), there is overall excellent agreement between theory and measurement. The frequency noise levels measured in Fig. \ref{fig2}b are consistent with fundamental SBL frequency noise (see Methods). Significantly, the fundamental nature of the noise, the good agreement between the PF prediction (Eq. (3)) and measurement in Fig. \ref{fig2}b, and separate experimental work \cite{Gyro} that has verified the theoretical form of the SEF (Eq. (2)) confirm that SEF $=$ PF so that the fundamental SNR of the gyroscope does not improve near the EP. 

While the Petermann factor analysis provides very good agreement with the measured results, we also derived an Adler-like coupled mode equation analysis for the Brillouin laser system. This approach is distinct from the bi-orthogonal framework and, while more complicated, provides additional insights into the system behavior. Adapting analysis applied in the noise analysis of ring laser gyroscopes \cite{AdlerNoise}, a noise enhancement factor NEF results (see Supplementary information),
\begin{equation}
\label{NEF}
\mathrm{NEF}=\frac{\Delta\omega_\mathrm{D}^2+\Delta\omega_\mathrm{EP}^2/2}{\Delta\omega_\mathrm{D}^2-\Delta\omega_\mathrm{EP}^2}
\end{equation}
It is interesting that this result, despite the different physical context of the Brillouin laser system, has a similar form to one derived for polarization-mode-coupled laser systems \cite{ExcessNoiseEP2}. The PF and NEF predictions are shown on Fig. \ref{fig2}b and the Adler-derived NEF correction provides slightly better agreement with the data at the intermediate detuning values.

\begin{figure}
\centering
\includegraphics[width=8.5cm]{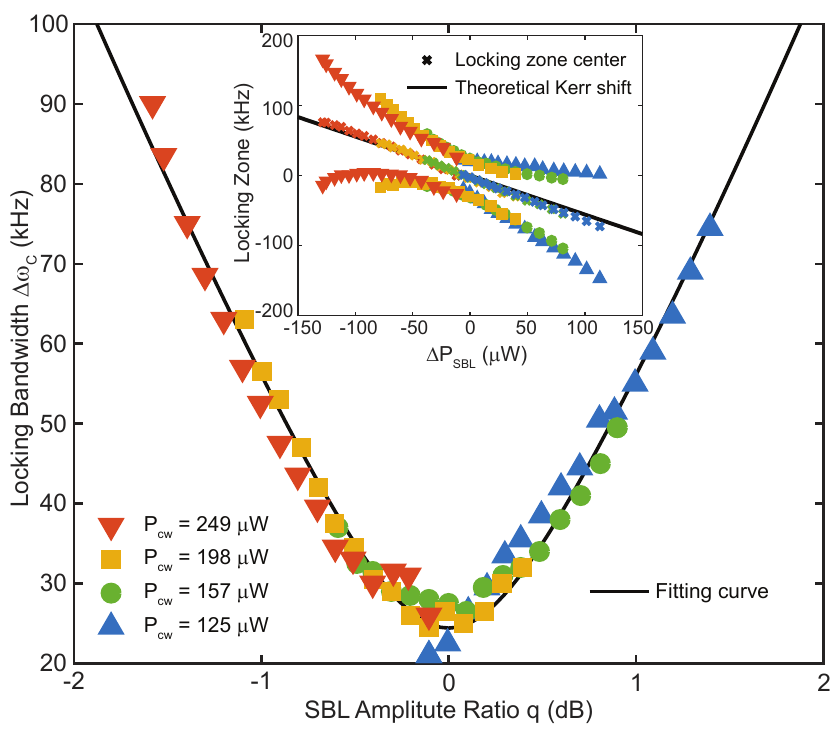}
\caption{\textbf{Locking zone bandwidth versus SBL amplitude ratio.}
Measured locking zone bandwidth is plotted versus amplitude ratio $q$ of the SBL lasers. The cw power is held constant at four values (see legend) to create the data composite. The solid black curve is Eq, (\ref{LockBand}). Inset: the measured locking zone boundaries are plotted versus the SBL power differences ($\Delta P_\mathrm{SBL}=P_\mathrm{ccw}-P_\mathrm{cw}$). Colors and symbols correspond to the main panel. The center of the locking zone is also indicated and is shifted by the Kerr nonlinearity which varies as the SBL power difference. Black line gives the theoretical prediction (no free parameters).}
\label{fig3}
\end{figure}

The Adler approach is also useful to explain a locking zone dependence upon SBL amplitudes observed in Fig. \ref{fig2}a. As shown in the Supplementary Information, this variation can be explained through the combined action of the Kerr effect and intermodal coupling coefficients of both dissipative and conservative nature. Specifically, the locking bandwidth is found to exhibit the following dependence upon the amplitude ratio $q=|a_\mathrm{ccw}/a_\mathrm{cw}|$ of the SBL lasers,
\begin{equation}
\label{LockBand}
\Delta\omega_\mathrm{EP}^2=\left(\frac{\Gamma}{\Gamma+\gamma}\right)^2\left[\left(q+\frac{1}{q}\right)^2|\kappa|^2+\left(q-\frac{1}{q}\right)^2|\chi|^2\right]
\end{equation}
where $\kappa$ is the dissipative coupling and $\chi$ is the conservative coupling between cw and ccw SBL modes. The locking zone boundaries in terms of pump detuning frequency have been measured (Fig. \ref{fig3} inset) for a series of different SBL powers. Using this data, the locking bandwidth is expressed in pump frequency detuning ($\Delta \omega_{\rm P}$) units using $\Delta \omega_C \equiv (1+\Gamma/\gamma) \Delta\omega_{\mathrm{EP}}$ and plotted versus $q$ in the main panel of Fig. \ref{fig3}. The plot agrees well with Eq. (\ref{LockBand}) (fitting shown in black) and gives $|\kappa|$ = 0.93 kHz, $|\chi|$ = 8.21 kHz.

Finally, the center of the locking band is shifted by the Kerr effect and (in pump frequency detuning $\Delta \omega_{\rm P}$ units) can be expressed as $-(\Gamma/\gamma)\Delta\omega_\mathrm{Kerr}$, where $\Delta\omega_\mathrm{Kerr}=
\eta\left(\left|\alpha_{\mathrm{ccw}}\right|^2-\left|\alpha_{\mathrm{cw}}\right|^2\right) = (\eta\Delta P_{\mathrm{SBL}})/(\gamma_\mathrm{ex}\hbar\omega)$ is the Kerr induced SBL resonance frequency difference, $\Delta P_\mathrm{SBL}=P_\mathrm{ccw}-P_\mathrm{cw}$ is the output power difference of the SBLs, and $\gamma_\mathrm{ex}$ is the photon decay rate due to the output coupling.
Also, $\eta=n_2\hbar\omega^2c/(Vn_0^2)$ is the single-photon Kerr-effect angular frequency shift with $\omega$ the SBL angular frequency, $n_2$ the Kerr-nonlinear refractive index of silica, $V$ the mode volume, $n_0$ the linear refractive index, and $c$ the speed of light in vacuum. If the white frequency noise floors in Fig. \ref{fig2} are used to infer the resonator quality factor, then a Kerr nonlinearity value of 558 Hz/{\textmu}W is predicted (see Methods). This value gives the line plot in the Fig. \ref{fig3} inset (with no free parameters) which agrees with experiment.

We have verified through measurement and theory that mode non-orthogonality sets a fundamental limitation to a class of sensors operating near an EP. Remarkably, a resulting noise enhancement precisely compensates the sensor's EP-enhanced response. It is nonetheless important to note that when SNR is limited by technical noise considerations, it still could be advantageous to operate near the EP. It is also possible that other sensing modalities could benefit from operation near an EP. More generally, the excellent control of the state space that is possible in the Brillouin system can provide a new platform for study of the remarkable physics associated with exceptional points.

\ 

\noindent {\bf Methods}

\noindent {\it Linewidth and Allan deviation measurement.}
In experiments, frequency is measured in the time domain using a frequency counter and its Allan deviation is calculated for different averaging times (Fig. \ref{fig2}a). 
The Allan deviation $\sigma_\nu(\tau)$ for a signal frequency is defined by
\begin{equation}
\sigma_\nu\left(\tau\right)\equiv\sqrt{\frac{1}{2\left(M-1\right)}\sum_{k=1}^{M-1}\left(\overline{\nu}_{k+1}-\overline{\nu}_{k}\right)^2}
\end{equation}
where $\tau$ is the averaging time, $M$ is the number of frequency measurements, and $\overline{\nu}_{k}$ is the average frequency of the signal (measured in Hz) in the time interval between $k\tau$ and $(k+1)\tau$.  The Allan deviation follows a $\tau^{-1/2}$ dependence when the underlying frequency noise spectral density is white \cite{IEEEstandard} as occurs for laser frequency noise limited by spontaneous emission. White noise causes the lineshape of the laser to be a Lorentzian. White noise is also typically dominant in the Allan deviation plot at shorter averaging times where flicker noise and frequency drift are not yet important. This portion of the Allan deviation plot can be fit using $\sigma_\nu(\tau)=\sqrt{S_\nu/(2\tau)}$ where $S_\nu$ is the white frequency noise one-sided spectral density function. This result can be further converted to the Lorentzian full-width at half maximum (FWHM) linewidth $\Delta\nu_\mathrm{FWHM}$ (measured in Hz) using the conversion,
\begin{equation}
S_\nu=2\sigma_\nu^2(\tau)\tau=\frac{1}{\pi}\Delta\nu_\mathrm{FWHM}
\end{equation}

\noindent {\it Experimental parameters and data fitting.}
The resonator is pumped at the optical wavelength $\lambda = 1553.3$ nm, which, subject to the Brillouin phase matching condition, corresponds to a phonon frequency (Stokes frequency shift) of approximately $\Omega_\mathrm{phonon}/(2\pi)= 10.8$ GHz. Quality factors of the SBL modes are measured using a Mach-Zehnder interferometer, and a loaded Q factor $Q_\mathrm{T} = 88\times10^6$ and coupling Q factor $Q_\mathrm{ex} = 507\times10^6$ are obtained. 

The theoretical formula for the white frequency noise of the beat frequency far away from the EP reads,
\begin{equation}
S_\nu=\left(\frac{\Gamma}{\gamma+\Gamma}\right)^2\frac{\hbar \omega^3}{4\pi^2Q_\mathrm{T}Q_\mathrm{ex}}(\frac{1}{P_\mathrm{cw}}+\frac{1}{P_\mathrm{ccw}})(n_\mathrm{th}+N_\mathrm{th}+1)
\end{equation}
which results from summing the Schawlow-Townes-like linewidths of the SBL laser waves \cite{SBL1}. In the expression, $N_\mathrm{th}$ and $n_\mathrm{th}$ are the thermal occupation numbers of the SBL state and phonon state, respectively. At room temperature, $n_\mathrm{th} \approx 577$ and $N_\mathrm{th} \approx 0$. For the power balanced case (green data set in Fig. \ref{fig2}), $P_\mathrm{cw} = P_\mathrm{ccw} = 215$ $\mu$W and the predicted white frequency noise (Eq. (8)) is $S_\nu$ = 0.50 Hz$^2$/Hz. For the blue (red) data set, $P_\mathrm{cw}$ ($P_\mathrm{ccw}$) is decreased by 1.22 dB (1.46 dB) so that $S_\nu =$ 0.58 (0.60) Hz$^2$/Hz is calculated. On the other hand, the measured values for the blue, green and red data sets in Fig. \ref{fig2}b (i.e., white frequency noise floors far from EP) give $S_\nu =$  0.44, 0.39, 0.46 Hz$^2$/Hz, respectively. The difference here is attributed to errors in Q measurement. For example, the experimental values of noise can be used to infer a corrected coupling Q factor $Q_{ex}\approx 658\times 10^6$. Using this value below yields an excellent prediction of the Kerr nonlinear coefficient which supports this belief.  

The beating frequency in Fig. \ref{fig2}a is fit using the following relations:
\begin{equation}
\begin{split}
\Delta\omega_\mathrm{S} & = \mathrm{sgn}(\Delta\omega_\mathrm{D})\sqrt{\Delta\omega_\mathrm{D}^2-\Delta\omega_\mathrm{EP}^2}\\
\Delta\omega_\mathrm{D} & = \frac{\gamma/\Gamma}{1+\gamma/\Gamma}\Delta\omega_\mathrm{P}+\frac{1}{1+\gamma/\Gamma}\Delta\omega_{\mathrm{Kerr}}
\end{split}
\end{equation}
where sgn is the sign function and $\gamma/\Gamma$, $\Delta\omega_{\mathrm{Kerr}}$ and $\Delta\omega_\mathrm{EP}$ are fitting parameters. The fitting gives $\gamma/\Gamma = 0.076$ consistently, while $\Delta\omega_{\mathrm{Kerr}}$ and $\Delta\omega_{\mathrm{EP}}$ are separately adjusted in each data set. These parameters feature a power dependence that is fully explored in Fig. \ref{fig3} and the related main text discussion.

The theoretical Kerr coefficient used in Fig. \ref{fig3} can be calculated as follows. Assuming $n_2\approx2.7\times10^{-20}$ $\text{m}^2/\text{W}$, $n_0=1.45$ for the silica material, and $V=10^7$ $\mu\text{m}^3$ (obtained through finite-element simulations for the 36mm-diameter disk used here), gives $\eta/2\pi\approx 10^{-5}$ Hz. Using the $Q_\mathrm{ex}$ corrected by the white frequency noise data (see discussion above), $\gamma_{ex}/2\pi = 299$ kHz so that $\Delta\omega_{\mathrm{Kerr}}/(2\pi\Delta P_{\mathrm{SBL}})\approx 42$ $\text{Hz}/\mu\text{W}$. When $\gamma/\Gamma=0.076$, the center shift of pump locking band is $-(\Gamma/\gamma)\Delta\omega_{\mathrm{Kerr}}$ = 558 Hz/$\mu$W.
This value agrees very well with experiment (Fig. \ref{fig3} inset).

\ 

\noindent {\it Data availability.} The data that support the plots within this paper and other findings of this study are available from the corresponding author upon reasonable request.

\

\noindent {\it Acknowledgements.} This work was supported by the Defense Advanced Research Projects Agency (DARPA) under PRIGM:AIMS program through SPAWAR (grant no. N66001-16-1-4046) and the Kavli Nanoscience Institute. 

\

\noindent {\it Author contributions.} HW, Y-HL and KV conceived the idea. HW derived the theory with the feedback from Y-HL, ZY and KV. Y-HL designed and perform the experiments with ZY and HW. ZY analysed the data with Y-HL and HW. M-GS fabricated the devices. All authors participated in writing the manuscript. KV supervised the research.

\

\noindent {\it Competing interests.} The authors declare no competing interests.

\

\noindent {\it Author Information.}
Current addresses of two co-authors: 
Y-HL, OEwaves Inc., 465 North Halstead Street, Suite 140, Pasadena, California 91107, USA;
M-GS, NTT Physics and Information Laboratory, 1950 University Ave., East Palo Alto, California 94303, USA.

\bibliography{main}

\clearpage
\onecolumngrid
\appendix
\renewcommand{\theequation}{S\arabic{equation}}
\renewcommand{\thefigure}{S\arabic{figure}}
\setcounter{figure}{0}
\setcounter{equation}{0}

\section{Supplementary information}
The supplementary information is structured as follows. In Section I we briefly review the framework for working with general non-Hermitian matrices and introduce the bi-orthogonal relations. In section II we derive the Petermann factor for a $2\times 2$ Hamiltonian. In Section III we show that the effective amplitude of the non-orthogonal modes is reduced compared to conventional modes and also justify the physical picture of the increased noise. Finally in section IV we present the full coupled mode-equations in a Langevin formalism. It is used to derive an Adler-like equation that will lead to the corrected noise enhancement factor as well as a relation giving the locking bandwidth as a function of the amplitude ratio of the two modes.

\section{Non-Hermitian Hamiltonian and bi-orthogonal relations}
Here we briefly review the framework for working with general non-Hermitian matrices. An $n$-dimensional matrix $M$ has $n$ eigenvalues $\mu_1,\ \mu_2,\ ...\ \mu_n$. For simplicity we will assume that all of the eigenvalues are distinct, i.e. $\mu_j\neq\mu_k$ if $j\neq k$. In this case $M$ will have $n$ right eigenvectors and $n$ left eigenvectors associated with each $\mu_j$:
\begin{equation}
M|v^\mathrm{R}_j\rangle=\mu_j|v^\mathrm{R}_j\rangle,\ \ \langle v^\mathrm{L}_j|M=\langle v^\mathrm{L}_j|\mu_j
\end{equation}
To make sense of the left eigenvectors, note that $M^\dag|v^\mathrm{L}_j\rangle=\mu_j^*|v^\mathrm{L}_j\rangle$, thus the left eigenvector is the eigenstate as if loss is changed to gain and vice versa. Since $M$ is in general non-Hermitian, there is no guarantee that $|v^\mathrm{L}_j\rangle=|v^\mathrm{R}_j\rangle$, and many of the decomposition results that hold in the Hermitian case will fail. However we note that,
\begin{equation}
\mu_j\langle v^\mathrm{L}_j|v^\mathrm{R}_k\rangle=\langle v^\mathrm{L}_j|M|v^\mathrm{R}_k\rangle=\mu_k\langle v^\mathrm{L}_j|v^\mathrm{R}_k\rangle\ \Rightarrow\ \langle v^\mathrm{L}_j|v^\mathrm{R}_k\rangle=0,\ \forall j\neq k
\end{equation}
Thus left and right eigenvectors associated with different eigenvalues are bi-orthogonal. We also note that the right eigenvectors are complete and form a set of basis (as $M$ is non-degenerate and finite-dimensional), and we can decompose the identity matrix and $M$ as follows:
\begin{equation}
\begin{split}
1 & =\sum_{j}\frac{|v^\mathrm{R}_j\rangle\langle v^\mathrm{L}_j|}{\langle v^\mathrm{L}_j|v^\mathrm{R}_j\rangle}\\
M & =\sum_{j}\frac{|v^\mathrm{R}_j\rangle\langle v^\mathrm{L}_j|}{\langle v^\mathrm{L}_j|v^\mathrm{R}_j\rangle}\mu_j
\end{split}
\end{equation}
where each term is a ``projector'' onto the eigenvectors. Again we note that $\langle v^\mathrm{L}_j|v^\mathrm{R}_j\rangle$ may be negative and even complex, which results in special normalizations of the vectors. For simplicity we will choose $\langle v^\mathrm{L}_j|v^\mathrm{R}_j\rangle=1$ by rescaling the vectors and adjusting the relative phase (such vectors are sometimes said to be bi-orthonormal). With this normalization in place the above decompositions simplify further as follows:
\begin{equation}
\begin{split}
1 & =\sum_{j}|v^\mathrm{R}_j\rangle\langle v^\mathrm{L}_j|\\
M & =\sum_{j}|v^\mathrm{R}_j\rangle\langle v^\mathrm{L}_j|\mu_j
\end{split}
\end{equation}
We note that, as a result of using bi-orthonormal left and right vectors, the vectors, themselves, are not normalized, i.e. $\langle v^\mathrm{L}_j|v^\mathrm{L}_j\rangle$ and $\langle v^\mathrm{R}_j|v^\mathrm{R}_j\rangle$ need not be $1$ for each $j$. There is one extra degree of freedom per mode for fixing the lengths, but the length normalization factors do not affect the physical observables if such factors are kept consistently through the calculations. In section III a ``natural'' normalization will be chosen when we give a physical meaning to these factors. 

\section{Petermann factor of a two-dimensional Hamiltonian}
In this section we derive the Petermann factor of a two-dimensional Hamiltonian $H$. Denote the two normalized right (left) eigenvectors of $H$ as $|\psi^\mathrm{R}_1\rangle$ and $|\psi^\mathrm{R}_2\rangle$ ($|\psi^\mathrm{L}_1\rangle$ and $|\psi^\mathrm{L}_2\rangle$). The Petermann factors of these two eigenmodes can then be expressed as \cite{PF2}
\begin{equation}
\begin{split}
\mathrm{PF}_1 & =\langle\psi^\mathrm{L}_1|\psi^\mathrm{L}_1\rangle\langle\psi^\mathrm{R}_1|\psi^\mathrm{R}_1\rangle \\
\mathrm{PF}_2 & =\langle\psi^\mathrm{L}_2|\psi^\mathrm{L}_2\rangle\langle\psi^\mathrm{R}_2|\psi^\mathrm{R}_2\rangle \\
\end{split}
\end{equation}
We will first prove that $\mathrm{PF}_1=\mathrm{PF}_2$, which can then be identified as the Petermann factor for the entire system. Note that $|\psi^\mathrm{L}_1\rangle$ and $|\psi^\mathrm{R}_2\rangle$ are orthogonal and span the two-dimensional space. As a result, the identity can be expressed using this set of basis vectors as follows:
\begin{equation}
1=\frac{|\psi^\mathrm{L}_1\rangle\langle \psi^\mathrm{L}_1|}{\langle \psi^\mathrm{L}_1|\psi^\mathrm{L}_1\rangle}+\frac{|\psi^\mathrm{R}_2\rangle\langle \psi^\mathrm{R}_2|}{\langle \psi^\mathrm{R}_2|\psi^\mathrm{R}_2\rangle}
\end{equation}
Now apply this expansion to $|\psi^\mathrm{R}_1\rangle$ and obtain
\begin{equation}
\label{S2R1expansion}
|\psi^\mathrm{R}_1\rangle=\frac{1}{\langle \psi^\mathrm{L}_1|\psi^\mathrm{L}_1\rangle}|\psi^\mathrm{L}_1\rangle+\frac{\langle \psi^\mathrm{R}_2|\psi^\mathrm{R}_1\rangle}{\langle \psi^\mathrm{R}_2|\psi^\mathrm{R}_2\rangle}|\psi^\mathrm{R}_2\rangle
\end{equation}
where $\langle\psi^\mathrm{L}_1|\psi^\mathrm{R}_1\rangle=1$ has been used. Left multiplication by $\langle\psi^\mathrm{R}_1|$ results in
\begin{equation}
\langle\psi^\mathrm{R}_1|\psi^\mathrm{R}_1\rangle=\frac{1}{\langle \psi^\mathrm{L}_1|\psi^\mathrm{L}_1\rangle}+\frac{\langle \psi^\mathrm{R}_1|\psi^\mathrm{R}_2\rangle\langle \psi^\mathrm{R}_2|\psi^\mathrm{R}_1\rangle}{\langle \psi^\mathrm{R}_2|\psi^\mathrm{R}_2\rangle}
\end{equation}
Thus we obtain,
\begin{equation}
\frac{1}{\mathrm{PF}_1}=1-\frac{\langle \psi^\mathrm{R}_1|\psi^\mathrm{R}_2\rangle\langle \psi^\mathrm{R}_2|\psi^\mathrm{R}_1\rangle}{\langle \psi^\mathrm{R}_1|\psi^\mathrm{R}_1\rangle\langle \psi^\mathrm{R}_2|\psi^\mathrm{R}_2\rangle}
\end{equation}
which is symmetric with respect to the indexes $1$ and $2$ and thereby completes the proof that $\mathrm{PF}_1=\mathrm{PF}_2\equiv \mathrm{PF}$.

Next, PF is expressed using the Hamiltonian instead of its eigenvectors. We begin by noting that the identity operator added to the Hamiltonian will not modify the eigenvectors. As a result, the trace can be removed from $H$ without changing the value of PF:
\begin{equation}
H_0\equiv H-\frac{1}{2}\mathrm{Tr}(H)
\end{equation}
where $\mathrm{Tr}$ is the matrix trace and $H_0$ is the traceless part of $H$. Using the bi-orthogonal expansion, $H_0$ has the form,
\begin{equation}
H_0=\mu(|\psi^\mathrm{R}_1\rangle\langle \psi^\mathrm{L}_1|-|\psi^\mathrm{R}_2\rangle\langle \psi^\mathrm{L}_2|)
\end{equation}
where $\mu$ is the first eigenvalue. Consider next the quantity $\mathrm{Tr}(H_0^\dag H_0)$:
\begin{equation}
\mathrm{Tr}(H_0^\dag H_0)=|\mu|^2(\langle\psi^\mathrm{L}_1|\psi^\mathrm{L}_1\rangle\langle\psi^\mathrm{R}_1|\psi^\mathrm{R}_1\rangle+\langle\psi^\mathrm{L}_2|\psi^\mathrm{L}_2\rangle\langle\psi^\mathrm{R}_2|\psi^\mathrm{R}_2\rangle-\langle\psi^\mathrm{L}_2|\psi^\mathrm{L}_1\rangle\langle\psi^\mathrm{R}_1|\psi^\mathrm{R}_2\rangle-\langle\psi^\mathrm{L}_1|\psi^\mathrm{L}_2\rangle\langle\psi^\mathrm{R}_2|\psi^\mathrm{R}_1\rangle)
\end{equation}
where we used the fact that $\mathrm{Tr}(|\alpha\rangle\langle\beta|)=\langle\beta|\alpha\rangle$. To simplify the expression, note that each of the first two terms equals PF. Moreover, the third term can be evaluated by expressing $|\psi^\mathrm{L}_1\rangle$ as a combination of right eigenvectors using Eq. (\ref{S2R1expansion}):
\begin{equation}
-\langle\psi^\mathrm{L}_2|\psi^\mathrm{L}_1\rangle\langle\psi^\mathrm{R}_1|\psi^\mathrm{R}_2\rangle=\frac{\langle \psi^\mathrm{R}_1|\psi^\mathrm{R}_2\rangle\langle \psi^\mathrm{R}_2|\psi^\mathrm{R}_1\rangle}{\langle \psi^\mathrm{R}_2|\psi^\mathrm{R}_2\rangle}\langle \psi^\mathrm{L}_1|\psi^\mathrm{L}_1\rangle=\mathrm{PF}-1
\end{equation}
Similarly, the fourth term also equals $\mathrm{PF}-1$. Thus
\begin{equation}
\mathrm{Tr}(H_0^\dag H_0)=|\mu|^2(4PF-2)
\end{equation}
Finally, to eliminate the eigenvalue $\mu$ we calculate,
\begin{equation}
\mathrm{Tr}(H_0^2)=\mu^2(\langle \psi^\mathrm{L}_1|\psi^\mathrm{R}_1\rangle^2+\langle \psi^\mathrm{L}_2|\psi^\mathrm{R}_2\rangle^2)=2\mu^2
\end{equation}
and the PF can be solved as
\begin{equation}
\mathrm{PF}=\frac{1}{2}\left(1+\frac{\mathrm{Tr}(H_0^\dag H_0)}{|\mathrm{Tr}(H_0^2)|}\right)
\end{equation}
which completes the proof.

We note that while a Hermitian Hamiltonian with $H_0^\dag=H_0$ results in $\mathrm{PF}=1$, the converse is not always true. Consider the example of $H_0=i\sigma_z$ where $\sigma_z$ is the Pauli matrix. This would effectively describe two orthogonal modes with different gain, and direct calculation shows that $\mathrm{PF}=1$.

\section{Field amplitude and noise in a non-orthogonal system}

In this section we consider the physical interpretation of increased linewidth whereby the effective field amplitude decreases while the effective noise input increases as a result of non-orthogonality. This analysis considers a hypothetical laser mode that is part of the bi-orthogonal system. It skips key steps normally taken in a more rigorous laser noise analysis in order to make clearer the essential EP physics. A more complete study of the Brillouin laser system is provided in Section IV. 

The two-dimensional system is described by the column vector $|\Psi\rangle=(a_\mathrm{cw},a_\mathrm{ccw})^T$ whose components are the orthogonal field amplitudes $a_\mathrm{cw}$ and $a_\mathrm{ccw}$. The equation of motion reads $id|\Psi\rangle/dt=H|\Psi\rangle$, where $H$ is the two-dimensional Hamiltonian. Now assume that $|\Psi\rangle=c_1|\psi^\mathrm{R}_1\rangle$,  i.e. only the first eigenmode of the system is excited. We interpret $c_1$ as the phasor for the eigenmode. We see that $|c_1|^2 = \langle\Psi|\Psi\rangle /  \langle\psi^\mathrm{R}_1|\psi^\mathrm{R}_1\rangle$ is reduced from the true square amplitude $\langle\Psi|\Psi\rangle$ by a factor of the length squared of the right eigenvector $\langle\psi^\mathrm{R}_1|\psi^\mathrm{R}_1\rangle$. The equation of motion for $c_1$ reads
\begin{equation}
i\frac{dc_1}{dt}=i\frac{d\langle\psi^\mathrm{L}_1|\Psi\rangle}{dt}=\langle\psi^\mathrm{L}_1|H_0|\psi^\mathrm{R}_1\rangle c_1=\mu_1 c_1
\end{equation}
Here, we are assuming that the mode experiences both loss and saturable gain that are absorbed into the definition of the eigenvalue $\mu_1$. To simplify the following calculations we set the real part of $\mu_1$ to $0$,  since any frequency shift can be removed with an appropriate transformation to slowly varying amplitudes. 

To introduce noise into the system resulting from the amplification process the equation of motion is modified as follows: $id|\Psi\rangle/dt=H_0|\Psi\rangle+|F\rangle$. Here, $|F\rangle=(\overline{F}_\mathrm{cw}(t),\overline{F}_\mathrm{ccw}(t))^T$ is a column vector with fluctuating components. The noise correlation of these components is assumed to be given by,
\begin{equation}
\label{S3corr}
\begin{split}
\langle \overline{F}_\mathrm{cw}^*(t)\overline{F}_\mathrm{cw}(t')\rangle = \langle \overline{F}_\mathrm{ccw}^*(t)\overline{F}_\mathrm{ccw}(t')\rangle & = \theta \delta(t-t')\\
\langle \overline{F}_\mathrm{cw}^*(t)\overline{F}_\mathrm{ccw}(t')\rangle = \langle \overline{F}_\mathrm{ccw}^*(t)\overline{F}_\mathrm{cw}(t')\rangle & = 0
\end{split}
\end{equation}
where $\theta$ is a quantity with frequency dimensions. We note that the assumption of vanishing correlations between the fluctuations on different modes is not trivial. Even if the basis is orthogonal, the non-Hermitian nature of the Hamiltonian means that dissipative mode coupling will generally be present in the system. This will be associated with fluctuations that can induce off-diagonal elements in the correlation matrix. In the system studied here, we will show in Section IV that the main source of noise comes from the phonons and fluctuations due to the non-Hermitian Hamiltonian are negligible, thereby justifying the assumption made here. Taking account of the fluctuations, the equation of motion for $c_1$ can be modified as follows,
\begin{equation}
\frac{dc_1}{dt}=-|\mu_1|c_1+\langle\psi^\mathrm{L}_1|F\rangle=-|\mu_1|c_1+\overline{F}_1
\end{equation}
where the fluctuation term for the first eigenmode is defined as $\overline{F}_1=\langle\psi^\mathrm{L}_1|F\rangle$. Its correlation reads
\begin{equation}
\langle \overline{F}_1^*(t)\overline{F}_1(t') \rangle =\theta \langle\psi^\mathrm{L}_1|\psi^\mathrm{L}_1\rangle \delta(t-t')
\end{equation}
which, upon comparison to Eq. (\ref{S3corr}), shows that the noise input to the right eigenvector field amplitude ($c_1$) is enhanced (relative to the noise input to either the cw or ccw fields alone) by a factor of the length squared of the left eigenvector $\langle\psi^\mathrm{L}_1|\psi^\mathrm{L}_1\rangle$. 

We are interested in the phase fluctuations of $c_1$. Here, it is assumed that the mode is pumped to above threshold and is lasing. Under these conditions it is possible separate amplitude and phase fluctuations of the field. We rewrite $c_1=|c_1|\exp(-i\phi_c)$ and obtain the rate of change of the phase variable as follows: 
\begin{equation}
\frac{d\phi_c}{dt}=\frac{i}{2|c_1|}\left(\overline{F}_1 e^{i\phi_c}-\overline{F}_1^* e^{-i\phi_c}\right)
\end{equation}
which describes white frequency noise of the laser field (equivalently phase noise diffusion). The correlation can be calculated as
\begin{equation}
\langle\dot{\phi}_c(t)\dot{\phi}_c(t')\rangle=\frac{\theta}{2|c_1|^2} \langle\psi^\mathrm{L}_1|\psi^\mathrm{L}_1\rangle\delta(t-t') =\frac{\theta}{2\langle\Psi|\Psi\rangle}\langle\psi^\mathrm{R}_1|\psi^\mathrm{R}_1\rangle\langle\psi^\mathrm{L}_1|\psi^\mathrm{L}_1\rangle\delta(t-t') = \mathrm{PF}\times\frac{\theta}{2\langle\Psi|\Psi\rangle}\delta(t-t')
\end{equation}
where the non-enhanced linewidth is $\Delta \omega_0 = \theta/(2\langle\Psi|\Psi\rangle)$ \cite{LaserPhysics} and the enhanced linewidth is given by $\Delta \omega = \mathrm{PF} \times \Delta \omega_0$. From the above derivation, the PF enhancement is the result of two effects, the reduction of effective square amplitude ($|c_1|^2 = \langle\Psi|\Psi\rangle /  \langle\psi^\mathrm{R}_1|\psi^\mathrm{R}_1\rangle$) and the enhancement of noise by $\langle\psi^\mathrm{L}_1|\psi^\mathrm{L}_1\rangle$. 

Up to now we have not chosen individual normalizations for $\langle\psi^\mathrm{L}_1|\psi^\mathrm{L}_1\rangle$ and $\langle\psi^\mathrm{R}_1|\psi^\mathrm{R}_1\rangle$ as they appear together in the Petermann factor. Motivated by the fact that left and right eigenvectors can be mapped onto the same Hilbert space, we select the symmetric normalization:
\begin{equation}
\langle\psi^\mathrm{L}_1|\psi^\mathrm{L}_1\rangle=\langle\psi^\mathrm{R}_1|\psi^\mathrm{R}_1\rangle=\sqrt{\mathrm{PF}}
\end{equation}
With this normalization the squared field amplitude is reduced and the noise input is increased by a factor of $\sqrt{\mathrm{PF}}$, resulting in the linewidth enhancement by a factor of PF. We note that other interpretations are possible through different normalizations. For example, in Siegman's analysis $\langle\psi^\mathrm{L}_1|\psi^\mathrm{L}_1\rangle=\mathrm{PF}$ and $\langle\psi^\mathrm{R}_1|\psi^\mathrm{R}_1\rangle=1$ is chosen, and the enhancement is fully attributed to noise increase by a factor of PF \cite{PF2} .

\section{Langevin formalism}

Here we analyze the system with a Langevin formalism. An Adler-like equation will be derived that provides an improved laser linewidth and and an expression for the locking bandwidth dependence on the field amplitude ratio. The analysis will also include the Kerr effect.

First we summarize symbols and give their definitions. For readability, all cw subscript will be replaced by $\overline{1}$ and all ccw subscript will be replaced by $\overline{2}$. The modes are pumped at angular frequencies $\omega_\mathrm{P,\overline{1}}$ and $\omega_\mathrm{P,\overline{2}}$. These frequencies will generally be different from the unpumped resonator mode frequency.  The cw and ccw Brillouin lasers oscillate on the same longitudinal mode with frequency $\omega$. This frequency is shifted for both cw and ccw waves by the same amount as a result of the pump-induced Kerr shift.  On the other hand, the Kerr effect causes cross-phase and self-phase modulation of the cw and ccw waves that induces different frequency shifts in these waves. This shift and the rotation-induced Sagnac shift are accounted for using offset frequencies $\delta\omega_{\overline{1}}=-\eta\left(a_{\overline{1}}^\dag a_{\overline{1}}+2a_{\overline{2}}^\dag a_{\overline{2}}\right)-\Omega \omega D/(2n_gc)$ and $\delta\omega_{\overline{2}}=-\eta\left(a_{\overline{2}}^\dag a_{\overline{2}}+2a_{\overline{1}}^\dag a_{\overline{1}}\right)+\Omega \omega D/(2n_gc)$ relative to  $\omega$, where $\eta=n_2\hbar\omega^2c/(Vn_0^2)$ is the single-photon nonlinear angular frequency shift, $n_2$ is the nonlinear refractive index, $V$ is the mode volume, $n_0$ is the linear refractive index, $c$ is the speed of light in vacuum, $\Omega$ is the rotation rate, $D$ is the resonator diameter and $n_g$ is the group index. Phonon modes have angular frequencies $\Omega_\mathrm{phonon}=2\omega n_0 v_s/c$ where $v_s$ is the velocity of the phonons. The loss rate of phonon modes is denoted as $\Gamma$ (also known as the gain bandwidth) and the loss rate of the SBL modes are assumed equal and denoted as $\gamma$. In addition, coupling between the two SBL modes is separated as a dissipative part and conservative part, denoted as $\kappa$ and $\chi$, respectively. These rates will be assumed to satisfy $\Gamma\gg\gamma\gg|\kappa|,|\chi|$ to simplify the calculations, which is {\it a posteriori} verified in our system. In the following analysis, we will treat the SBL modes and phonon modes quantum mechanically and define $a_{\overline{1}}$ ($a_{\overline{2}}$) and $b_{\overline{1}}$ ($b_{\overline{2}}$) as the lowering operators of the cw (ccw) components of the SBL and phonon modes, respectively. Meanwhile, pump modes are treated as a noise-free classical fields $A_{\overline{1}}$ and $A_{\overline{2}}$ (photon-number-normalized amplitudes).

Using these definitions, the full equations of motion for the SBL and phonon modes read
\begin{equation}
\begin{split}
\dot{a}_{\overline{1}} & = -\left(\frac{\gamma}{2}+i\omega +i\delta\omega_{\overline{1}}\right) a_{\overline{1}} + (\kappa+i\chi) a_{\overline{2}}-i g_{ab} A_{\overline{2}} b_{\overline{2}}^\dag\exp(-i\omega_\mathrm{P,\overline{2}}t)+F_{\overline{1}}(t)\\
\dot{a}_{\overline{2}} & = -\left(\frac{\gamma}{2}+i\omega +i\delta\omega_{\overline{2}}\right) a_{\overline{2}} + (\kappa^*+i\chi^*) a_{\overline{1}}-i g_{ab} A_{\overline{1}} b_{\overline{1}}^\dag\exp(-i\omega_\mathrm{P,\overline{1}}t)+F_{\overline{2}}(t)\\
\dot{b}_{\overline{1}}^\dag & = -\left(\frac{\Gamma}{2}-i\Omega_\mathrm{phonon}\right) b_{\overline{1}}^\dag + ig_{ab} A_{\overline{1}}^* a_{\overline{2}} \exp(i\omega_\mathrm{P,\overline{1}}t)+ f_{\overline{1}}^\dag(t)\\
\dot{b}_{\overline{2}}^\dag & = -\left(\frac{\Gamma}{2}-i\Omega_\mathrm{phonon}\right) b_{\overline{2}}^\dag + ig_{ab} A_{\overline{2}}^* a_{\overline{1}} \exp(i\omega_\mathrm{P,\overline{2}}t)+ f_{\overline{2}}^\dag(t)
\end{split}
\end{equation}where $g_{ab}$ is the single-particle Brillioun coupling coefficient. The fluctuation operators $F(t)$ and $f(t)$ associated with the field operators have the following correlations:
\begin{equation}
\begin{split}
\langle F_{\overline{1}}^\dag(t)F_{\overline{1}}(t') \rangle = \langle F_{\overline{2}}^\dag(t)F_{\overline{2}}(t') \rangle & = \gamma N_\mathrm{th} \delta(t-t')\\
\langle F_{\overline{1}}(t)F_{\overline{1}}^\dag(t') \rangle = \langle F_{\overline{2}}(t)F_{\overline{2}}^\dag(t') \rangle & = \gamma (N_\mathrm{th}+1) \delta(t-t')\\
\langle f_{\overline{1}}^\dag(t)f_{\overline{1}}(t') \rangle = \langle f_{\overline{2}}^\dag(t)f_{\overline{2}}(t') \rangle & = \Gamma n_\mathrm{th} \delta(t-t')\\
\langle f_{\overline{1}}(t)f_{\overline{1}}^\dag(t') \rangle = \langle f_{\overline{2}}(t)f_{\overline{2}}^\dag(t') \rangle & = \Gamma (n_\mathrm{th}+1) \delta(t-t')
\end{split}
\end{equation}
where $N_\mathrm{th}$ and $n_\mathrm{th}$ are the thermal occupation numbers of the SBL state and phonon state. In addition, there are non-zero cross-correlations of the photon fluctuation operators due to the dissipative coupling:
\begin{equation}
\begin{split}
\langle F_{\overline{2}}^\dag(t)F_{\overline{1}}(t') \rangle & = -2\kappa N_\mathrm{th} \delta(t-t')\\
\langle F_{\overline{1}}^\dag(t)F_{\overline{2}}(t') \rangle & = -2\kappa^* N_\mathrm{th} \delta(t-t')\\
\langle F_{\overline{2}}(t)F_{\overline{1}}^\dag(t') \rangle & = -2\kappa^* (N_\mathrm{th}+1) \delta(t-t')\\
\langle F_{\overline{1}}(t)F_{\overline{2}}^\dag(t') \rangle & = -2\kappa (N_\mathrm{th}+1) \delta(t-t')\\
\end{split}
\end{equation}All other cross correlations not explicitly written are $0$.

\subsection{Single SBL}

We first study a single laser mode and its corresponding phonon field  ($a_{\overline{1}}$ and $b_{\overline{2}}$) by neglecting $\kappa$ and $\chi$. By introducing the slow varying envelope with $a_{\overline{1}}=\alpha_{\overline{1}} e^{-i\omega t}$ and $b_{\overline{2}}=\beta_{\overline{2}} e^{-i(\omega_{\mathrm{P},\overline{2}}-\omega) t}$, the following equations result:
\begin{equation}
\begin{split}
\dot{\alpha}_{\overline{1}} & = -\left(\frac{\gamma}{2} +i\delta\omega_{\overline{1}}\right) \alpha_{\overline{1}} -i g_{ab} A_{\overline{2}} \beta_{\overline{2}}^\dag+F_{\overline{1}}(t)e^{i\omega t}\\
\dot{\beta}_{\overline{2}}^\dag & = -\left(\frac{\Gamma}{2}+i\Delta\Omega_{\overline{2}}\right) \beta_{\overline{2}}^\dag + ig_{ab} A_{\overline{2}}^* \alpha_{\overline{1}}+ f_{\overline{2}}^\dag(t)e^{-i(\omega_\mathrm{P,\overline{2}}-\omega) t}
\end{split}
\end{equation}
where we have defined the frequency mismatch $\Delta\Omega_{\overline{2}}=\omega_{\mathrm{P},\overline{2}}-\omega-\Omega_\mathrm{phonon}$.  Neglecing the weak Kerr effect term in $\delta\omega_{\overline{1}}$ this is a set of linear equations in $a_{\overline{1}}$ and $b_{\overline{2}}$. The eigenvalues of the coefficient matrix
\begin{equation*}
\left(
\begin{matrix}
-\gamma/2-i\delta\omega_{\overline{1}} & -i g_{ab}A_{\overline{2}}\\
i g_{ab} A_{\overline{2}}^* & -\Gamma/2-i\Delta\Omega_{\overline{2}}
\end{matrix}
\right)
\end{equation*}
can be solved as
\begin{equation}
\mu_{1,2}=\frac{1}{4}\left(-\Gamma-\gamma-2i\delta\omega_{\overline{1}}-2i\Delta\Omega_{\overline{2}}\pm\sqrt{16g_{ab}^2|A_{\overline{2}}|^2+(\Gamma-\gamma-2i\delta\omega_{\overline{1}}+2i\Delta\Omega_{\overline{2}})^2}\right)
\end{equation}
At the lasing threshold, the first eigenvalue $\mu_1$ has a real part of 0. This can be rewritten as
\begin{equation}
16g_{ab}^2|A_{\overline{2}}|^2+(\Gamma-\gamma-2i\delta\omega_{\overline{1}}+2i\Delta\Omega_{\overline{2}})^2=(\Gamma+\gamma+2i\delta\omega_{\overline{1}}+2i\Delta\Omega_{\overline{2}}+4i\mathrm{Im}(\mu_1))^2
\end{equation}
Solving this complex equation gives the SBL eigenfrequency as well as the lasing threshold,
\begin{equation}
\mu_1=-i\frac{\gamma\Delta\Omega_{\overline{2}}+\Gamma\delta\omega_{\overline{1}}}{\Gamma+\gamma}
\end{equation}
\begin{equation}
g_{ab}^2|A_{\overline{2}}|^2=\frac{\Gamma\gamma}{4}\left(1+\frac{4(\Delta\Omega_{\overline{2}}-\delta\omega_{\overline{1}})^2}{(\Gamma+\gamma)^2}\right)
\end{equation}
The threshold at perfect phase matching ($\Delta\Omega_{\overline{2}}=\delta\omega_{\overline{1}}$) is usually written in a more familiar form $g_0|A_{\overline{2}}|^2=\gamma/2$, where $g_0$ is the Brillouin gain factor \cite{SBL1}. Comparison gives
\begin{equation}
g_{ab}=\sqrt{\frac{g_0\Gamma}{2}}
\end{equation}
We also introduce the modal Brillioun gain function for a single direction:
\begin{equation}
g_{\overline{1}} \equiv \frac{g_0}{1+4(\delta\omega_{\overline{1}}-\Delta\Omega_{\overline{2}})^2/(\Gamma+\gamma)^2}
\end{equation}
so that the threshold can be written as
\begin{equation}
g_{\overline{1}}|A_{\overline{2}}|^2=\gamma/2
\end{equation}

With the threshold condition solved, the matrix can be decomposed using the bi-orthogonal approach outlined in Section I. The linear combination that describes the composite SBL mode can be found as
\begin{equation}
\overline{\alpha}_{\overline{1}}=\frac{\Gamma}{\gamma+\Gamma}\left[\alpha_{\overline{1}}-i\frac{g_{ab}}{\Gamma}\frac{2}{1+2i(\Delta\Omega_{\overline{2}}-\delta\omega_{\overline{1}})/(\Gamma+\gamma)}A_{\overline{2}}\beta_{\overline{2}}^\dag\right]
\end{equation}
where the factor $\Gamma/(\gamma+\Gamma)$ properly normalizes $\overline{\alpha}_{\overline{1}}$ so that $\overline{\alpha}_{\overline{1}}=\alpha_{\overline{1}}$ when only the SBL mode is present in the system, and we have dropped its dependence on the phase mismatch $\Delta\Omega_{\overline{2}}-\delta\omega_{\overline{1}}$ for simplicity. The associated equation of motion is
\begin{equation}
\frac{d}{dt}\overline{\alpha}_{\overline{1}}=-i\frac{\gamma\Delta\Omega_{\overline{2}}+\Gamma\delta\omega_{\overline{1}}}{\Gamma+\gamma}\overline{\alpha}_{\overline{1}}+\overline{F}_{\overline{1}}(t)
\end{equation}
where the frequency term now includes a mode-pulling contribution so that the SBL laser frequency is given by, 
\begin{equation}
\omega_\mathrm{S,\overline{1}}=\omega+\frac{\gamma\Delta\Omega_{\overline{2}}+\Gamma\delta\omega_{\overline{1}}}{\Gamma+\gamma}
\end{equation}
Also, we have defined a combined fluctuation operator for $\overline{\alpha}_{\overline{1}}$,
\begin{equation}
\overline{F}_{\overline{1}}(t)=\frac{\Gamma}{\gamma+\Gamma}\left[F_{\overline{1}}(t)e^{i\omega t}-i\sqrt{\frac{1-2i(\Delta\Omega_{\overline{2}}-\delta\omega_{\overline{1}})/(\Gamma+\gamma)}{1+2i(\Delta\Omega_{\overline{2}}-\delta\omega_{\overline{1}})/(\Gamma+\gamma)}}\sqrt{\frac{\gamma}{\Gamma}}f_{\overline{2}}^\dag(t)e^{-i(\omega_\mathrm{P,\overline{2}}-\omega) t}\right]
\end{equation}
with the following correlations,
\begin{equation}
\begin{split}
\langle \overline{F}_{\overline{1}}^\dag(t)\overline{F}_{\overline{1}}(t') \rangle & = \left(\frac{\Gamma}{\gamma+\Gamma}\right)^2\left(\langle F_{\overline{1}}^\dag(t)F_{\overline{1}}(t') \rangle + \frac{\gamma}{\Gamma} \langle f_{\overline{1}}^\dag(t)f_{\overline{1}}(t') \rangle \right) = \left(\frac{\Gamma}{\gamma+\Gamma}\right)^2 \gamma (n_\mathrm{th}+N_\mathrm{th}) \delta(t-t')\\
\langle \overline{F}_{\overline{1}}(t)\overline{F}_{\overline{1}}^\dag(t') \rangle & = \left(\frac{\Gamma}{\gamma+\Gamma}\right)^2\left(\langle F_{\overline{1}}(t)F_{\overline{1}}^\dag(t') \rangle + \frac{\gamma}{\Gamma} \langle f_{\overline{1}}(t)f_{\overline{1}}^\dag(t') \rangle \right) = \left(\frac{\Gamma}{\gamma+\Gamma}\right)^2 \gamma (n_\mathrm{th}+N_\mathrm{th}+2) \delta(t-t')
\end{split}
\end{equation}

We can now write $\overline{\alpha}_{\overline{1}}(t)=\sqrt{N_{\overline{1}}}\exp(-i\phi_{\overline{1}})$, where $N_{\overline{1}}$ is the photon number, $\phi_{\overline{1}}$ is the phase for the SBL, and where amplitude fluctuations have been ignored on account of quenching of these fluctuations above laser threshold. We note that amplitude fluctuations may result in linewidth corrections similar to the Henry $\alpha$ factor, but we will ignore these effects here. The full equation of motion for $\phi_{\overline{1}}$ is
\begin{equation}
\dot{\phi}_{\overline{1}}=\omega_\mathrm{S,\overline{1}}-\omega+\Phi_{\overline{1}}(t),\ \ \Phi_{\overline{1}}(t)=\frac{i}{2\sqrt{N_{\overline{1}}}}(\overline{F}_{\overline{1}}(t)e^{i\phi_{\overline{1}}}-\overline{F}_{\overline{1}}^\dag(t)e^{-i\phi_{\overline{1}}})
\end{equation}
The correlation of the noise operator is given by,
\begin{equation}
\langle \Phi_{\overline{1}}(t) \Phi_{\overline{1}}(t') \rangle=\frac{1}{4N_{\overline{1}}}(\langle \overline{F}_{\overline{1}}^\dag(t)\overline{F}_{\overline{1}}(t') \rangle+\langle \overline{F}_{\overline{1}}(t)\overline{F}_{\overline{1}}^\dag(t') \rangle=\left(\frac{\Gamma}{\gamma+\Gamma}\right)^2\frac{\gamma}{2N_{\overline{1}}}(n_\mathrm{th}+N_\mathrm{th}+1)\delta(t-t')
\end{equation}
and we identify the coefficient before the delta function,
\begin{equation}
\Delta\omega_\mathrm{FWHM,\overline{1}}=\left(\frac{\Gamma}{\gamma+\Gamma}\right)^2\frac{\gamma}{2N_{\overline{1}}}(n_\mathrm{th}+N_\mathrm{th}+1)
\end{equation}
as the full-width half-maximum (FWHM) linewidth of the SBL. 

In the experiment, the frequency noise of the SBL beating signal is measured. To compare against the experiment, we calculate the FWHM linewidth for the beating signal by adding together the linewidths in two directions:
\begin{equation}
\Delta\omega_\mathrm{FWHM}=\Delta\omega_\mathrm{FWHM,\overline{1}}+\Delta\omega_\mathrm{FWHM,\overline{2}}=\left(\frac{\Gamma}{\gamma+\Gamma}\right)^2\left(\frac{1}{2N_{\overline{1}}}+\frac{1}{2N_{\overline{2}}}\right)\gamma(n_\mathrm{th}+N_\mathrm{th}+1)
\end{equation}
and then convert to the one-sided power spectral density $S_\nu$:
\begin{equation}
S_\nu=\frac{1}{\pi}\frac{\Delta\omega_\mathrm{FWHM}}{2\pi}=\left(\frac{\Gamma}{\gamma+\Gamma}\right)^2\frac{\hbar \omega^3}{4\pi^2Q_\mathrm{T}Q_\mathrm{ex}}(\frac{1}{P_\mathrm{cw}}+\frac{1}{P_\mathrm{ccw}})(n_\mathrm{th}+N_\mathrm{th}+1)
\end{equation}
where $Q_\mathrm{T}$ and $Q_\mathrm{ex}$ are the loaded and coupling $Q$ factors, and $P_\mathrm{cw}$ and $P_\mathrm{ccw}$ are the SBL powers in each direction.

\subsection{Two SBLs}

Now we can apply a similar procedure to the two pairs of photon and phonon modes with coupling on the optical modes. We write the equations of motion for the SBL modes:
\begin{equation}
\begin{split}
\frac{d}{dt}\overline{\alpha}_{\overline{1}} &= -i(\omega_\mathrm{S,\overline{1}}-\omega)\overline{\alpha}_{\overline{1}}+\frac{\Gamma}{\gamma+\Gamma}(\kappa+i\chi)\alpha_{\overline{2}}+\overline{F}_{\overline{1}}(t)\\
\frac{d}{dt}\overline{\alpha}_{\overline{2}} &= -i(\omega_\mathrm{S,\overline{2}}-\omega)\overline{\alpha}_{\overline{2}}+\frac{\Gamma}{\gamma+\Gamma}(\kappa^*+i\chi^*)\alpha_{\overline{1}}+\overline{F}_{\overline{2}}(t)
\end{split}
\end{equation}
where quantities with the opposite subscript are defined similarly to those in the previous section. We note that the coupling term involves the optical modes $\alpha_{\overline{1}}$ and $\alpha_{\overline{2}}$ only. However, no additional coupling occurs between the components of the SBL eigenststes $\overline{\alpha}_{\overline{1}}$ and $\overline{\alpha}_{\overline{2}}$, and these states do not change up to first order of $\kappa/\gamma$. Thus we can approximate the optical mode $\alpha_{\overline{1}}$ with the SBL mode $\overline{\alpha}_{\overline{1}}$. With these approximations the lasing thresholds are also the same as the independent case \cite{Gyro}. The equations now become
\begin{equation}
\begin{split}
\frac{d}{dt}\overline{\alpha}_{\overline{1}} &= -i(\omega_\mathrm{S,\overline{1}}-\omega)\overline{\alpha}_{\overline{1}}+(\overline{\kappa}+i\overline{\chi})\overline{\alpha}_{\overline{2}}+\overline{F}_{\overline{1}}(t)\\
\frac{d}{dt}\overline{\alpha}_{\overline{2}} &= -i(\omega_\mathrm{S,\overline{2}}-\omega)\overline{\alpha}_{\overline{2}}+(\overline{\kappa}^*+i\overline{\chi}^*)\overline{\alpha}_{\overline{1}}+\overline{F}_{\overline{2}}(t)
\end{split}
\end{equation}
where we have defined mode-pulled coupling rates $\overline{\kappa}=\kappa\Gamma/(\gamma+\Gamma)$ and $\overline{\chi}=\chi\Gamma/(\gamma+\Gamma)$. 

We can write $\overline{\alpha}_j(t)=\sqrt{N_j}\exp(-i\phi_j)$ with $j=\overline{1},\overline{2}$, and once again ignore amplitude fluctuations. The equations of motion for the phases are,
\begin{equation}
\begin{split}
\frac{d}{dt}\phi_{\overline{1}} & = (\omega_\mathrm{S,\overline{1}}-\omega)-q\mathrm{Im}[(\overline{\kappa}+i\overline{\chi})e^{(i\phi_{\overline{1}}-i\phi_{\overline{2}})}]+\frac{i}{2\sqrt{N_{\overline{1}}}}(\overline{F}_{\overline{1}}(t)e^{i\phi_{\overline{1}}}-\overline{F}_{\overline{1}}^\dag(t)e^{-i\phi_{\overline{1}}})\\
\frac{d}{dt}\phi_{\overline{2}} & = (\omega_\mathrm{S,\overline{2}}-\omega)-q^{-1}\mathrm{Im}[(\overline{\kappa}^*+i\overline{\chi}^*)e^{(i\phi_{\overline{2}}-i\phi_{\overline{1}})}]+\frac{i}{2\sqrt{N_{\overline{2}}}}(\overline{F}_{\overline{2}}(t)e^{i\phi_{\overline{2}}}-\overline{F}_{\overline{2}}^\dag(t)e^{-i\phi_{\overline{2}}})\\
\end{split}
\end{equation}
where we have defined the amplitude ratio $q=\sqrt{N_{\overline{2}}/N_{\overline{1}}}$ for simplicity. As we measure the beatnote frequency, it is convenient to define $\phi\equiv\phi_{\overline{2}}-\phi_{\overline{1}}$ from which we obtain,
\begin{equation}
\frac{d\phi}{dt} = (\omega_\mathrm{S,\overline{2}}-\omega_\mathrm{S,\overline{1}})+\mathrm{Im}\left\{\left[q(\overline{\kappa}+i\overline{\chi})+q^{-1}(\overline{\kappa}-i\overline{\chi})\right]e^{-i\phi}\right\}+\Phi(t)\\
\end{equation}
where the combined noise term and its correlation are given by
\begin{equation}
\Phi =-\frac{i}{2\sqrt{N_{\overline{1}}}}(\overline{F}_{\overline{1}}(t)e^{i\phi_{\overline{1}}}-\overline{F}_{\overline{1}}^\dag(t)e^{-i\phi_{\overline{1}}}) +\frac{i}{2\sqrt{N_{\overline{2}}}}(\overline{F}_{\overline{2}}(t)e^{i\phi_{\overline{2}}}-\overline{F}_{\overline{2}}^\dag(t)e^{-i\phi_{\overline{2}}})
\end{equation}
\begin{equation}
\langle \Phi(t) \Phi(t') \rangle =\left(\frac{\Gamma}{\gamma+\Gamma}\right)^2 \left[\left(\frac{1}{2N_{\overline{1}}}+\frac{1}{2N_{\overline{2}}}\right)\gamma(N_\mathrm{th}+n_\mathrm{th}+1)+\frac{2}{\sqrt{N_{\overline{1}}N_{\overline{2}}}}\left(N_\mathrm{th}+\frac{1}{2}\right)\mathrm{Re}(\kappa e^{-i\phi(t)})\right]\delta(t-t')
\end{equation}
Since both $N_\mathrm{th}$ and $\kappa/\gamma$ is small, we will discard the last time-varying term and write
\begin{equation}
\langle \Phi(t) \Phi(t') \rangle \approx \Delta\omega_\mathrm{FWHM}\delta(t-t')
\end{equation}
where $\Delta\omega_\mathrm{FWHM}=\Gamma^2/(\gamma+\Gamma)^2[(2N_{\overline{1}})^{-1}+(2N_{\overline{2}})^{-1}]\gamma(N_\mathrm{th}+n_\mathrm{th}+1)$ is the linewidth of the beating signal far from the EP (see also the single SBL discussion).

This equation can be further simplified by introducing an overall phase shift with $\overline{\phi}=\phi-\phi_0$, where $\phi_0=\mathrm{Arg}\left[q(\overline{\kappa}+i\overline{\chi})+q^{-1}(\overline{\kappa}-i\overline{\chi})\right]$ and $\mathrm{Arg}(z)$ is the phase of $z$:
\begin{equation}
\frac{d\overline{\phi}}{dt} = \Delta\omega_\mathrm{D}-\Delta\omega_\mathrm{EP}\sin\overline{\phi}+\Phi(t)
\end{equation}
with
\begin{equation}
\Delta\omega_\mathrm{D}=\omega_\mathrm{S,\overline{2}}-\omega_\mathrm{S,\overline{1}}=\frac{\gamma}{\Gamma+\gamma}(\omega_\mathrm{P,\overline{1}}-\omega_\mathrm{P,\overline{2}})+\frac{\Gamma}{\Gamma+\gamma}\left[\eta(N_{\overline{2}}-N_{\overline{1}})+\frac{\omega D}{n_gc}\Omega\right]
\end{equation}
\begin{equation}
\Delta\omega_\mathrm{EP}^2=\left|q(\overline{\kappa}+i\overline{\chi})+q^{-1}(\overline{\kappa}-i\overline{\chi})\right|^2=\left(\frac{\Gamma}{\gamma+\Gamma}\right)^2\left[\left(q+\frac{1}{q}\right)^2|\kappa|^2+\left(q-\frac{1}{q}\right)^2|\chi|^2+2\left(q^2-\frac{1}{q^2}\right)\mathrm{Im}(\kappa\chi^*)\right]
\end{equation}
This is an Adler equation with a noisy input. It shows the dependence of locking bandwidth on the amplitude ratio and coupling coefficients. Moreover, it is clear that in the absence of $\Delta\omega_\mathrm{EP}$, the beating linewidth would be given by $\Delta\omega_\mathrm{FWHM}$. The locking term $\Delta\omega_\mathrm{EP}\sin\overline{\phi}$ makes the rate of phase change nonuniform and increases the linewidth.

To analyze this stochastic Adler equation and obtain the linewidth, we define $z_\phi=\exp(-i\overline\phi)$ and rewrite
\begin{equation}
\frac{d}{dt}z_\phi=-iz_\phi(\Delta\omega_\mathrm{D}+\Delta\omega_\mathrm{EP}\frac{z_\phi-z_\phi^{-1}}{2i}+\Phi)
\end{equation}
The solution to the Adler equation is periodic when no noise is present. To see this explicitly we use a linear fractional transform:
\begin{equation}
z_t=\frac{(\Delta\omega_\mathrm{D}-\Delta\omega_\mathrm{S})z_\phi+i\Delta\omega_\mathrm{EP}}{\Delta\omega_\mathrm{EP} z_\phi+i(\Delta\omega_\mathrm{D}-\Delta\omega_\mathrm{S})},\ \ z_\phi=-i\frac{(\Delta\omega_\mathrm{D}-\Delta\omega_\mathrm{S})z_t-\Delta\omega_\mathrm{EP}}{\Delta\omega_\mathrm{EP}z_t-(\Delta\omega_\mathrm{D}-\Delta\omega_\mathrm{S})},\ \ 
|z_\phi|=|z_t|=1
\end{equation}
\begin{equation}
\frac{1}{z_t}\frac{d}{dt}z_t=i\Delta\omega_\mathrm{S}-i\frac{\Delta\omega_\mathrm{EP}(z_t+z_t^{-1})/2-\Delta\omega_\mathrm{D}}{\Delta\omega_\mathrm{S}}\Phi
\end{equation}
where we introduced $\Delta\omega_\mathrm{S}= \sqrt{\Delta\omega_\mathrm{D}^2-\Delta\omega_\mathrm{EP}^2}$ (which has the same meaning in the main text). The noiseless solution of $z_t$ would be $z_t=\exp(i\Delta\omega_\mathrm{S} t)$, and $z_\phi$ can be expanded in $z_t$ as
\begin{equation}
z_\phi=-i\frac{\Delta\omega_\mathrm{D}-\Delta\omega_\mathrm{S}}{\Delta\omega_\mathrm{EP}}+2i\frac{\Delta\omega_\mathrm{S}}{\Delta\omega_\mathrm{EP}}\sum_{p=1}^\infty\left(\frac{\Delta\omega_\mathrm{D}-\Delta\omega_\mathrm{S}}{\Delta\omega_\mathrm{EP}z_t}\right)^p
\end{equation}
where we have assumed $\Delta\omega_\mathrm{D}>\Delta\omega_\mathrm{EP}$ for convenience so that convergence can be guaranteed (For the case $\Delta\omega_\mathrm{D}<-\Delta\omega_\mathrm{EP}$ we can expand near $z_t=0$ instead of $z_t=\infty$). Thus the signal consists of harmonics oscillating at frequency $p\Delta\omega_\mathrm{S}$ with exponentially decreasing amplitudes. The noise added only changes the phase of $z_t$ (as the coefficient is purely imaginary) and to the lowest order the only effect of noise is to broaden each harmonic.

The linewidth can be found from the spectral density, which is given by the Fourier transform of the correlation function:
\begin{equation}
W_E(\omega)\propto \mathcal{F}_\tau\{\langle z^*_\phi(t) z_\phi(t+\tau) \rangle\}(\omega)
\end{equation}
and the correlation is given by
\begin{equation}
\langle z^*_\phi(t) z_\phi(t+\tau) \rangle =\left(\frac{\Delta\omega_\mathrm{D}-\Delta\omega_\mathrm{S}}{\Delta\omega_\mathrm{EP}}\right)^2+4\frac{\Delta\omega_\mathrm{S}^2}{\Delta\omega_\mathrm{EP}^2}\sum_{p=1}^\infty\left(\frac{\Delta\omega_\mathrm{D}-\Delta\omega_\mathrm{S}}{\Delta\omega_\mathrm{EP}}\right)^{2p}\langle z_t(t)^p z_t(t+\tau)^{-p}\rangle
\end{equation}
where we have discarded the $\langle z_t(t)^p z_t(t+\tau)^{-q} \rangle \ (p\neq q)$ terms since they vanish at the lowest order of $\Delta\omega_\mathrm{FWHM}$.

To further calculate each $\langle z_t(t)^p z_t(t+\tau)^{-p}\rangle \equiv C_p(\tau)$ we require the integral form of the Fokker-Planck equation: if $dX(t)=\mu(X,t)dt+\sigma(X,t)dW$ is a stochastic differential equation (in the Stratonovich interpretation), where $W$ is a Wiener process, then for $f(X)$ as a function of $X$, the differential equation for its average reads,
\begin{equation*}
\frac{d}{dt}\langle f(X) \rangle = \langle (\mu+\frac{\sigma}{2}\frac{\partial \sigma}{\partial X}) f'(X) \rangle + \langle \frac{1}{2} \sigma^2 f''(X) \rangle
\end{equation*}

Applying the Fokker-Planck equation to $C_p(\tau)$, with the stochastic equation for $z_t$, gives
\begin{equation}
\begin{split}
\frac{d C_p(\tau)}{d \tau} & = -p\langle i\Delta\omega_\mathrm{S} z_t(t)^p z_t(t+\tau)^{-p}\rangle \\
& +p\langle \left[\frac{\Delta\omega_\mathrm{FWHM}}{2\Delta\omega_\mathrm{S}^2}\left(\Delta\omega_\mathrm{EP}\frac{z_t(t+\tau)+z_t^{-1}(t+\tau)}{2}-\Delta\omega_\mathrm{D} \right)\left(\Delta\omega_\mathrm{EP}z_t(t+\tau)-\Delta\omega_\mathrm{D}\right)\right] z_t(t)^p z_t(t+\tau)^{-p}\rangle \\
& - p(p+1)\langle \frac{\Delta\omega_\mathrm{FWHM}}{2\Delta\omega_\mathrm{S}^2}\left(\Delta\omega_\mathrm{EP}\frac{z_t(t+\tau)+z_t^{-1}(t+\tau)}{2}-\Delta\omega_\mathrm{D} \right)^2 z_t(t)^p z_t(t+\tau)^{-p} \rangle \\
& \approx \left(-i p \Delta\omega_\mathrm{S} -p^2\frac{\Delta\omega_\mathrm{D}^2+\Delta\omega_\mathrm{EP}^2/2}{2\Delta\omega_\mathrm{S}^2}\Delta\omega_\mathrm{FWHM}\right)C_p(\tau)
\end{split}
\end{equation}
and $C_p(0)=1$, where again $\langle z_t(t)^p z_t(t+\tau)^{-q} \rangle \ (p\neq q)$ terms are discarded. Thus completing the Fourier transform for each term gives the linewidth of the respective harmonics. In particular, the linewidth of the fundamental frequency can be found through
\begin{equation}
W_{E,1}(\omega) \propto\frac{\Delta\overline{\omega}_\mathrm{FWHM}}{(\omega-\Delta\omega_\mathrm{S})^2+\Delta\overline{\omega}_\mathrm{FWHM}^2/4}
\end{equation}
with
\begin{equation}
\Delta\overline{\omega}_\mathrm{FWHM}=\frac{\Delta\omega_\mathrm{D}^2+\Delta\omega_\mathrm{EP}^2/2}{\Delta\omega_\mathrm{S}^2}\Delta\omega_\mathrm{FWHM}=\frac{\Delta\omega_\mathrm{D}^2+\Delta\omega_\mathrm{EP}^2/2}{\Delta\omega_\mathrm{D}^2-\Delta\omega_\mathrm{EP}^2}\Delta\omega_\mathrm{FWHM}
\end{equation}
We see that this result is different from the Petermann factor result, which is a theory linear in photon numbers and does not correctly take account of the saturation of the lasers and the Adler mode-locking effect.

From the expressions of $\Delta\omega_\mathrm{D}$ and $\Delta\omega_\mathrm{EP}$, the beating frequency can be expressed using the following hierarchy of equations:
\begin{equation}
\begin{split}
\Delta\omega_\mathrm{S} & = \mathrm{sgn}(\Delta\omega_\mathrm{D})\sqrt{\Delta\omega_\mathrm{D}^2-\Delta\omega_\mathrm{EP}^2}\\
\Delta\omega_\mathrm{D} & = \frac{\gamma}{\Gamma+\gamma}\Delta\omega_\mathrm{P}+\frac{\Gamma}{\Gamma+\gamma}\Delta\omega_\mathrm{Kerr}\\
\Delta\omega_\mathrm{P} & =\omega_\mathrm{P,\overline{1}}-\omega_\mathrm{P,\overline{2}}\\
\Delta\omega_\mathrm{Kerr} & = \eta(N_{\overline{2}}-N_{\overline{1}}) = \frac{\eta\Delta P_{\textrm{SBL}}}{\gamma_{ex}\hbar\omega}
\end{split}
\end{equation}
where $\mathrm{sgn}$ is the sign function and we take $\Omega=0$ (no rotation). For the Kerr shift,  $\Delta P_\mathrm{SBL}=P_\mathrm{ccw}-P_\textrm{cw}$ is the output power difference of the SBLs, and $\gamma_{ex}$ is the photon decay rate due to the output coupling. The center of the locking band can be found by setting $\Delta\omega_\mathrm{D}=0$, which leads to $\Delta\omega_\mathrm{P}=-(\Gamma/\gamma)\Delta\omega_\mathrm{Kerr}$.

We would like to remark that the equation for locking bandwidth $\Delta\omega_\mathrm{EP}$ in the main text
does not contain the phase-sensitive term $\mathrm{Im}(\kappa\chi^*)$. This terms leads to asymmetry of the locking band with respect to $q$ and $1/q$ and has not been observed in the experimental data.  We believe its contribution can be neglected. In other special cases, $\mathrm{Im}(\kappa\chi^*)$ disappears if there is a dominant, symmetric scatterer that determines both $\kappa$ and $\chi$ (e.g. the taper coupling point), or becomes negligible if there are many small scatterers that add up incoherently (e.g. surface roughness). This term can also be absorbed into the first two terms so the locking bandwidth is rewritten using effective $\kappa$, $\chi$ and a net amplitude imbalance $q_0$. Thus power calibration errors in the experiment can be confused with the phase-sensitive term in the locking bandwidth.

\end{document}